\documentclass[fleqn,12pt,twoside]{article}
\usepackage{espcrc1}

\usepackage {epsfig}

\usepackage{graphicx}
\usepackage[figuresright]{rotating}

\title{First Measurements of Pion Correlations by the PHENIX
Experiment}

\author{S.~C. Johnson\address{Physics and Advanced Technologies Directorate\\
Lawrence Livermore National Laboratory\\ 7000 East Ave. L-050 \\
Livermore, CA  94550,  USA}, for the PHENIX Collaboration \footnote{
For the full PHENIX Collaboration author list and acknowledgements see 
the contribution by W.A. Zajc (K. Adcox {\it et al.}) in this volume.}}

\begin{document}

\maketitle

\begin{abstract}
First identical-pion correlations measured at RHIC energies by PHENIX
are presented.  Two analyses with separate detectors, systematics, and
statistics provide consistent results.  The resulting HBT radii are
moderately larger than those measured at lower energies.
The $k_t$ dependence of
the Bertsch-Pratt HBT radii is also similar to previous measures and
is consistent with the conjecture of an expanding source.
\end{abstract}

\section{INTRODUCTION}

One predicted result of a first order phase
transition to a quark gluon plasma in heavy ion collisions is a long
lived state due to the large latent heat and corresponding reduction
in pressure gradients in the created system \cite{rischke}.  The primary method
for measuring 
the space-time extent of heavy-ion collisions is two particle
interferometry, which has been utilized in hadronic interactions
from proton induced to nuclear collisions. For an
incoherent source ($S$) of identical bosons,
the normalized probability of detecting two particles with 
relative momentum $q=p_1-p_2$ and average momentum $k=(p_1+p_2)/2$ is
given by the correlation function ($C_2$):

\begin{equation}
C_2(\vec{q},\vec{k}) =  1+{|\int dx S(x,\vec{k}) e^{iqx}|}^2
\label{eq:c2}
\end{equation}

\noindent
where $x$ is the four-position, and the integral is taken over all
space-time. 
The direct utilization of this technique is hindered in
principle because equation \ref{eq:c2} is uninvertible: the source has 7
independent variables while the measured correlation function has 6.
Furthermore, due to limited statistics most recent heavy ion
experiments have chosen to plot and fit the correlation function in
only three of the six dimensions versus the Bertsch-Pratt \cite{bpratt}
projection of $\vec{q}$ in the out-side-long directions:

\begin{equation}
C_2({\vec{q}},{\vec{k}}) = 1+ \lambda
\exp(-{q_{Tout}}^2{R_{Tout}}^2({\vec{k}})
     -{q_{Tside}}^2{R_{Tside}}^2({\vec{k}})
     -{q_{Long}}^2{R_{Long}}^2({\vec{k}}))
\end{equation}

\noindent
where $q_{Long}$ is the momentum difference in the beam direction,
$q_{Tout}$, in the pair momentum direction, and $q_{Tside}$, the
corresponding orthogonal direction.  A benefit
of such a fit is that, for a well-behaved, gaussian source with no
position-momentum correlations or resonances, the lifetime is simply
determined by the difference between $R_{Tside}$ and $R_{Tout}$:
$\beta^2\tau^2 = {R_{Tout}}^2-{R_{Tside}}^2$ \cite{fields}.  In such
studies the k-dependence of the HBT radii is usually explored by
repeating the fitting procedure with a subsample of the entire dataset
of pairs.

\section{PHENIX DETECTOR AND ANALYSIS}

The PHENIX experiment is described in detail
elsewhere \cite{morrison}.  For the analysis described in this article
we use the data collected in the summer of
2000.  After all offline analysis cuts, the data sample was
approximately 1.5 million events.

\begin{figure}
\begin{minipage}[b]{0.6\linewidth}
\mbox{\epsfig{figure=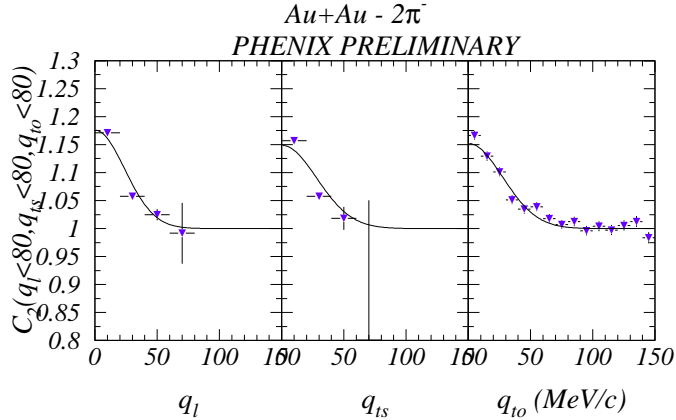,angle=0,height=5.9cm}}
\end{minipage}
\hfill
\parbox[b]{.3\textwidth}{\sloppy
\caption{Correlation function for $\pi^-$ pairs from the EMC
analysis.  The one-dimensional projections of the three-dimensional
correlation function is averaged over the lowest 80 MeV/c in the other
momentum differences.}
 \protect\label{fig:memc}}
\end{figure}

This analysis used a subset of all of the detectors in the
experiment.  For vertex information and centrality definitions we
rely on a pair of Cerenkov beam counters and zero-degree
calorimeters; the beam counters also act as the start of the time of
flight measurements.  The two drift chambers measure the deflection of
particles through the magnetic field and, hence, their momentum, while
the velocity is measured by either the time-of-flight wall (TOF) or the
electromagnetic calorimeter (EMC).  The TOF timing resolution is
approximately 115 ps \cite{phenix_pid}, while the
EMC resolution is 700 ps.  As a result, the TOF analysis provides
$\pi$-$K$ separation to 1.5 GeV/c; the EMC analysis, to 0.7 GeV/c.  Due to
geometric and acceptance effects the TOF analysis consists of $\sim
500,000$ ($\sim300,000$) pairs of identified $\pi^+$ ($\pi^-$) while the
EMC analysis consists of approximately five times as many pairs.

The particle identification algorithm is similar in both analyses.  A
pion is defined as being within 2$\sigma$ of the pion mass-squared peak
but 3$\sigma$ away from the kaon peak.  Backgrounds from long lived
resonances, e.g. $\Lambda$'s, are reduced by intradetector association cuts.
Energy deposition cuts in the TOF slat or EMC tower further reduce
backgrounds due to accidental misassociations.
Systematic studies have shown that the background contamination to the
EMC analysis is at least double that in the TOF analysis leading to
artificially lower $\lambda$ values in the former.  Ongoing
studies of the backgrounds in the analyses aim to resolve these
differences.  However, the introduction
of background from accidental hits or electron contamination tend to
only affect the resulting $\lambda$ in the fits and do not change the
measured radii.

A number of systematic studies have been performed to ensure that the
resulting correlation function does not include any artificial,
experimentally induced correlations from inefficiencies in either the
detector or tracking algoithms.  To remove these inefficiencies, pairs
of particles within 2 cm of each other in the drift chamber are
removed.  Further, pairs that share the same TOF slat or EMC cluster
are also removed in both the signal and mixed background.
For the Coulomb correction, the source is parameterized as a
gaussian in R$_{\rm inv}$=
$\sqrt{\delta r^2-\delta t^2}$, as 
determined by an iterative procedure \cite{baker}.

\section{RESULTS}
{
\begin{table}
\begin{tabular}{||c||c|c|c|c||}\hline
Data Set & $R_{Tout}$ (fm) & $R_{Tside}$ (fm) & $R_{Long}$ (fm) & $\lambda$ \\ \hline
\hline
EMC $\pi^+\pi^+$ & $4.4 \pm 0.2$ & $5.1 \pm 0.6$ & $5.9 \pm 0.4$ & $0.27
\pm .02$ \\ \hline
TOF $\pi^+\pi^+$ & $6.2 \pm 0.5$ & $7.9 \pm 1.1$ & $4.0 \pm 1.2$ & $0.49
\pm .05$ \\ \hline
EMC $\pi^-\pi^-$ & $5.1 \pm 0.2$ & $5.0 \pm 0.6$ & $5.9 \pm 0.4$ & $0.30
\pm .02$ \\ \hline
TOF $\pi^-\pi^-$ & $5.5 \pm 0.5$ & $5.8 \pm 1.5$ & $6.7 \pm 0.9$ & $0.49
\pm .06$ \\ \hline
\hline
\end{tabular}
\caption{Results of the Bertsch-Pratt fits to the identical pion pairs
in the EMC and TOF analyses.  Errors shown represent statistical
uncertainties only; current systematic uncertainties are $<1$ fm.}
\label{tab:results}
\end{table}
}

Figure \ref{fig:memc} shows the Coulomb corrected
correlation function for $\pi^{-}$ pairs measured in
the EMC analysis overlayed with the resulting fit.  While the fit is
performed in the full three dimensional space, we plot here projections of the
correlation function into each of the standard momentum difference
variables ($q_{Tout}$, $q_{Tside}$, and $q_{Long}$).  The results of
the fit are shown in Table \ref{tab:results} along with the results
from the $\pi^{+}$ analysis from the EMC and the $\pi^+$/$\pi^-$ analyses
utilizing the TOF.  The mean transverse momentum ($\langle k_t
\rangle$) of the pairs is 350 MeV/c
in the TOF analysis and 340  MeV/c in the EMC analysis
while the rapidity coverage in both analyses is centered about
mid-rapidity, $|y|<.35$.  The mean centrality of all pairs in the
analysis is $15$\% and is strongly biased towards central collisions.

Within the current statistical and systematic error bars, the EMC and
TOF analyses for both pion sets are consistent with one another.
The results of the fits are moderately larger than identical
measurements for Au+Au and Pb+Pb collisions at lower energies and
comparable $\langle k_t \rangle$ \cite{previous}.  
The results do not indicate an especially large source compared to
measurements at lower energies and the resulting naively calculated lifetime
deduced from these measurements is consistent with zero.  However, a
stronger collective flow will shorten the effective source
and lifetime measured by HBT.

\begin{figure}
\begin{minipage}[b]{0.6\linewidth}
\mbox{\epsfig{figure=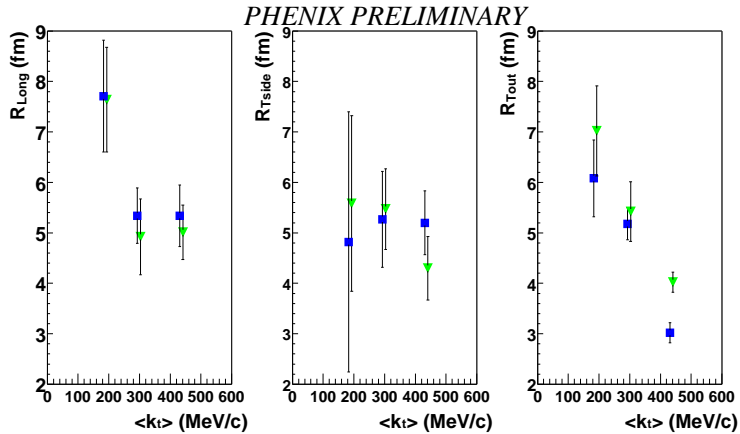,angle=0,height=6.0cm}}
\end{minipage}
\hfill
\parbox[b]{.3\textwidth}{\sloppy
\caption{$k_t$ dependence of the Bertsch-Pratt radii for $\pi^-$
(triangles) and $\pi^+$ (squares) pairs in the EMC analysis.  Error
bars correspond to statistical errors only.}
 \protect\label{fig:kt}}
\end{figure}

Plotted in Fig.~\ref{fig:kt} is the $\langle k_t \rangle$ dependence of the
radii in the EMC $\pi^+$ (squares) and $\pi^-$ (triangles)
analyses.  The sample is split into three approximately equal subsets
of pairs and the correlation analysis is performed on each subset.
The $k_t$ bins are $k_t < 250$ MeV/c, $250<k_t<350$ MeV/c and $k_t>350$ MeV/c
corresponding to $\langle k_t \rangle = 188$ MeV/c, $\langle k_t
\rangle = 298$ MeV/c and $\langle k_t \rangle = 
436$ MeV/c respectively.  Data points at identical $\langle k_t \rangle$
are offset by $10$ MeV/c for clarification.

The radii follow trends familiar from lower energy measurements at lower
beam energies \cite{previous}:  $R_{Tside}$ has a mild, if
any, dependence on the transverse momentum of the pair, while
$R_{Tout}$ varies strongly with $\langle k_t \rangle$.  Such
dependencies have been described by collective motions,
though temperature gradients and decays of short-lived resonances
could also contribute to such dependencies \cite{heinz_jacak}.

\section{CONCLUSION}

We have shown first measurements of identical pion correlations
measured by PHENIX at RHIC.  The resulting Bertsch-Pratt radii are
moderately larger than those measured at lower energies.  The $k_t$
dependence of the radii is consistent with the conjecture of a system
with a large degree of collective motion.

An extension of these measurements over an wider range in $k_t$ with
a larger data sample will be available following
the upcoming 2001 data collection.  During this second run the
additional PHENIX acceptance and integrated luminosity should lead to
approximately a factor of a hundred in pair statistics.  The resulting data
sample will lead to much stricter constraints on models of the
collision and allow for comparisons to identical kaon and proton HBT analyses.

This work was performed under the auspices of the U.S. Department of
Energy by the University of California, Lawrence Livermore National
Laboratory under Contract No. W-7405-Eng-48.


\begin{thebibliography}{9}
\bibitem{rischke} D.~H.~Rischke and M.~ Gyulassy, Nucl. Phys. A 
{\bf 608} (1996) 479.
\bibitem{bpratt} S. Pratt, Phys. Rev. D {\bf 33} (1986) 1314; G. Bertsch,
Phys. Rev. C {\bf 40} (1989) 1830.
\bibitem{fields} D.~E.~Fields, {\em et al.}, Phys. Rev. C {\bf 52} (1995)
986.
\bibitem{morrison} D.~P. Morrison for the PHENIX Collaboration,
Nucl. Phys. A {\bf 638} (1998) 565c.
\bibitem{phenix_pid} H.~Hamagaki for the PHENIX Collaboration, these
proceedings.
\bibitem{baker} M.~Baker, Nucl. Phus. A {\bf 610} (1996) 213c.


\bibitem{previous} I.~Bearden, {\em et al.} (NA44 Collaboration),
Phys. Rev. C {\bf 58} (1998) 1656; R.~A. Soltz, M.~Baker, J.~H.~Lee,
Nucl. Phys. A {\bf 661} (1999) 439c; R.~Ganz for the NA49
Collaboration, Nucl. Phus. {\bf A661} (1999) 448c; M.~A.~Lisa, {\em et
al.} (E895 Collaboration), Phys. Rev. Lett. {\bf 84} (2000) 2798.

\bibitem{heinz_jacak} U.~Heinz and B.~V.~Jacak,
Annu. Rev. Nucl. Part. Sci {\bf 49} (1999) 529.


\end{thebibliography}
\end{document}